\input harvmac

\def\fun#1#2{\lower3.6pt\vbox{\baselineskip0pt\lineskip.9pt
  \ialign{$\mathsurround=0pt#1\hfil##\hfil$\crcr#2\crcr\sim\crcr}}}
\relax


\noblackbox

\def\IZ{\relax\ifmmode\mathchoice
{\hbox{\cmss Z\kern-.4em Z}}{\hbox{\cmss Z\kern-.4em Z}} {\lower.9pt\hbox{\cmsss Z\kern-.4em Z}}
{\lower1.2pt\hbox{\cmsss Z\kern-.4em Z}}\else{\cmss Z\kern-.4em Z}\fi}
\def\IB{\relax{\rm I\kern-.18em B}}
\def\IC{{\relax\hbox{\kern.3em{\cmss I}$\kern-.4em{\rm C}$}}}
\def\ID{\relax{\rm I\kern-.18em D}}
\def\IE{\relax{\rm I\kern-.18em E}}
\def\IF{\relax{\rm I\kern-.18em F}}
\def\IG{\relax\hbox{$\inbar\kern-.3em{\rm G}$}}
\def\IGa{\relax\hbox{${\rm I}\kern-.18em\Gamma$}}
\def\IH{\relax{\rm I\kern-.18em H}}
\def\II{\relax{\rm I\kern-.18em I}}
\def\IK{\relax{\rm I\kern-.18em K}}
\def\IP{\relax{\rm I\kern-.18em P}}

\font\cmss=cmss10 \font\cmsss=cmss10 at 7pt
\def\IR{\relax{\rm I\kern-.18em R}}

\def\frac#1#2{{#1 \over #2}}

\def\OL#1{ \kern1pt\overline{\kern-1pt#1
   \kern-1pt}\kern1pt }

\lref\GVW{S. Gukov, C. Vafa and E. Witten, ``CFTs from Calabi-Yau Fourfolds,'' Nucl. Phys. {\bf B584} (2000) 69,
hep-th/9906070\semi T.R. Taylor and C. Vafa, ``RR flux on Calabi-Yau and partial supersymmetry breaking,'' Phys.
Lett. {\bf B474} (2000) 130, hep-th/9912152\semi P. Mayr, ``On Supersymmetry Breaking in String Theory and its
Realization in Brane Worlds,'' Nucl. Phys. {\bf B593} (2001) 99, hep-th/0003198.}

\lref\oldflux{ A. Strominger, ``Superstrings with Torsion,'' Nucl. Phys. {\bf B274} (1986) 253\semi J.
Polchinski and A. Strominger, ``New vacua for type II string theory,'' Phys. Lett. {\bf B388} (1996) 736,
hep-th/9510227\semi K. Becker and M. Becker, ``M-theory on eight manifolds,'' Nucl. Phys. {\bf B477} (1996) 155,
hep-th/9605053\semi J.Michelson, ``Compactifications of type IIB strings to four-dimensions with non-trivial
classical potential,'' Nucl. Phys. {\bf B495} (1997) 127, hep-th/9610151\semi K. Dasgupta, G. Rajesh and S.
Sethi, ``M-theory, orientifolds and G-flux,'' JHEP {\bf 9908} (1999) 023, hep-th/9908088\semi B. Greene, K.
Schalm and G. Shiu, ``Warped compactifications in M and F theory,'' Nucl. Phys. {\bf B584} (2000) 480,
hep-th/0004103\semi G. Curio, A. Klemm, D. L\"ust and S. Theisen, ``On the Vacuum Structure of Type II String
Compactifications on Calabi-Yau Spaces with H-Fluxes,'' Nucl. Phys. {\bf B609} (2001) 3, hep-th/0012213\semi K.
Becker and M. Becker, ``Supersymmetry Breaking, M Theory and Fluxes,'' JHEP {\bf 010} (2001) 038,
hep-th/0107044\semi M. Haack and J. Louis, ``M theory compactified on Calabi-Yau fourfolds with background
flux,'' Phys. Lett. {\bf B507} (2001) 296, hep-th/0103068\semi J. Louis and A. Micu, ``Type II theories
compactified on Calabi-Yau threefolds in the presence of background fluxes,'' Nucl.Phys. {\bf B635} (2002) 395,
hep-th/0202168.}

\lref\AdSorb{
S.~Kachru and E.~Silverstein, ``4d conformal theories and strings on orbifolds,'' Phys.\ Rev.\ Lett.\ {\bf 80},
4855 (1998) [arXiv:hep-th/9802183].
 }

\lref\Heterotic{ M. Becker, G. Curio and A. Krause, ``De Sitter Vacua from Heterotic M Theory,''
hep-th/0403027\semi R. Brustein and S.P. de Alwis, ``Moduli Potentials in String Compactifications with Fluxes:
Mapping the Discretuum,'' hep-th/0402088\semi S. Gukov, S. Kachru, X. Liu and L. McAllister, ``Heterotic Moduli
Stabilization with Fractional Chern-Simons Invariants,'' hep-th/0310159\semi E. Buchbinder and B. Ovrut,
``Vacuum Stability in Heterotic M-theory,'' hep-th/0310112.}

\lref\othermodnotes{
 M.~Dine, ``TASI lectures on M theory phenomenology,'' arXiv:hep-th/0003175;
S.~M.~Carroll,
arXiv:hep-th/0011110;
 }

\lref\simeon{S. Hellerman, ``On the Landscape of Superstring Theory in $D > 10$", hep-th/0405041}

\lref\karch{ A.~Karch, ``Auto-localization in de-Sitter space,'' JHEP {\bf 0307}, 050 (2003)
[arXiv:hep-th/0305192];
Lecture at Texas A.M. Conference on ``Strings and Cosmology",
http://mitchell.physics.tamu.edu/Conference/Cosmology04/Talks/Karch/karch.html. }

\lref\garyoliver{ O.~DeWolfe, D.~Z.~Freedman, S.~S.~Gubser, G.~T.~Horowitz and I.~Mitra, ``Stability of AdS(p) x
M(q) compactifications without supersymmetry,'' Phys.\ Rev.\ D {\bf 65}, 064033 (2002) [arXiv:hep-th/0105047].
}

\lref\dsww{
 M.~Dine, N.~Seiberg, X.~G.~Wen and E.~Witten, ``Nonperturbative Effects On The String World Sheet,'' Nucl.\
Phys.\ B {\bf 278}, 769 (1986);
M.~Dine, N.~Seiberg, X.~G.~Wen and E.~Witten,
Nucl.\ Phys.\ B {\bf 289}, 319 (1987).
}

\lref\instegs{
 F.~Denef, M.~R.~Douglas and B.~Florea, ``Building a Better Racetrack,'' arXiv:hep-th/0404257.
}

\lref\trapping{
L.~Kofman, A.~Linde, X.~Liu, A.~Maloney, L.~McAllister and E.~Silverstein, ``Beauty is attractive: Moduli
trapping at enhanced symmetry points,'' arXiv:hep-th/0403001.
}

\lref\AdStach{
A.~Adams and E.~Silverstein, ``Closed string tachyons, AdS/CFT, and large N QCD,'' Phys.\ Rev.\ D {\bf 64},
086001 (2001) [arXiv:hep-th/0103220].
}

\lref\IgorRR{
I.~R.~Klebanov, ``Tachyon stabilization in the AdS/CFT correspondence,'' Phys.\ Lett.\ B {\bf 466}, 166 (1999)
[arXiv:hep-th/9906220].
}

\lref\tongtach{
D.~Tong, ``Comments on condensates in non-supersymmetric orbifold field theories,'' JHEP {\bf 0303}, 022 (2003)
[arXiv:hep-th/0212235].
}

\lref\dinesymm{
M.~Dine, Y.~Nir and Y.~Shadmi, ``Enhanced symmetries and the ground state of string theory,'' Phys.\ Lett.\ B
{\bf 438}, 61 (1998) [arXiv:hep-th/9806124].
}

\lref\clusteringvac{
F.~Denef and M.~R.~Douglas, ``Distributions of flux vacua,'' arXiv:hep-th/0404116;
A.~Giryavets, S.~Kachru and P.~K.~Tripathy, ``On the Taxonomy of Flux Vacua,'' arXiv:hep-th/0404243.
}

\lref\kklmmt{
S.~Kachru, R.~Kallosh, A.~Linde, J.~Maldacena, L.~McAllister and S.~P.~Trivedi, ``Towards inflation in string
theory,'' JCAP {\bf 0310}, 013 (2003) [arXiv:hep-th/0308055].
}

\lref\KL{
D.~Kabat and G.~Lifschytz, ``Gauge theory origins of supergravity causal structure,'' JHEP {\bf 9905}, 005
(1999) [arXiv:hep-th/9902073].
}

\lref\Dccel{
E.~Silverstein and D.~Tong, ``Scalar speed limits and cosmology: Acceleration from D-cceleration,''
arXiv:hep-th/0310221;
M.~Alishahiha, E.~Silverstein and D.~Tong, ``DBI in the sky,'' arXiv:hep-th/0404084.
}

\lref\kls{L.~Kofman, A.~D.~Linde and A.~A.~Starobinsky, ``Towards the theory of reheating after inflation,''
Phys.\ Rev.\ D {\bf 56}, 3258 (1997), hep-ph/9704452.
}

\lref\HW{
 A.~Hanany and E.~Witten, ``Type IIB superstrings, BPS monopoles, and three-dimensional gauge
Nucl.\ Phys.\ B {\bf 492}, 152 (1997) [arXiv:hep-th/9611230].
 }

\lref\AdSdict{
E.~Witten, ``Anti-de Sitter space and holography,'' Adv.\ Theor.\ Math.\ Phys.\  {\bf 2}, 253 (1998)
[arXiv:hep-th/9802150].
S.~S.~Gubser, I.~R.~Klebanov and A.~M.~Polyakov, ``Gauge theory correlators from non-critical string theory,''
Phys.\ Lett.\ B {\bf 428}, 105 (1998) [arXiv:hep-th/9802109].
 }

\lref\junctions{
E.~Silverstein, ``AdS and dS entropy from string junctions,'' contribution to the Ian Kogan Memorial Volume,
arXiv:hep-th/0308175.
 }

\lref\vijayreview{
V.~Balasubramanian, ``Accelerating universes and string theory,'' Class.\ Quant.\ Grav.\ {\bf 21}, S1337 (2004)
[arXiv:hep-th/0404075].
 }

\lref\renata{
C.~P.~Burgess, R.~Kallosh and F.~Quevedo, ``de Sitter string vacua from supersymmetric D-terms,'' JHEP {\bf
0310}, 056 (2003) [arXiv:hep-th/0309187].
 }

\lref\gkpetc{ S. Giddings, S. Kachru, and J. Polchinski, ``Hierarchies from Fluxes in String
Compactifications,'' Phys. Rev. {\bf D66} (2002) 106006, hep-th/0105097; S. Kachru,~M. Schulz and S. P. Trivedi,
``Moduli Stabilization from Fluxes in a Simple IIB Orientifold,'' JHEP {\bf 0310} (2003) 007,
hep-th/0201028\semi A. Frey and J. Polchinski, ``N=3 Warped Compactifications,'' Phys. Rev. {\bf D65} (2002)
126009, hep-th/0201029; P. Tripathy and S. P. Trivedi, ``Compactification with Flux on K3 and Tori,''  JHEP {\bf
0303} (2003) 028, hep-th/0301139; A. Giryavets, S. Kachru, P. Tripathy and S. Trivedi, ``Flux Compactifications
on Calabi-Yau Threefolds,'' JHEP {\bf 0404} (2004) 003, hep-th/0312104.
 }

\lref\KPV{ S. Kachru, J. Pearson and H. Verlinde, ``Brane/Flux Annihilation and the String Dual of a
Non-supersymmetric Field Theory,'' JHEP {\bf 0206} (2002) 021, hep-th/0112197.}

\lref\joeI{
J.~Polchinski, ``String Theory. Vol. 1: An Introduction To The Bosonic String,''
 }

\lref\joeII{
J.~Polchinski, ``String Theory. Vol. 2: Superstring Theory And Beyond,''
 }

\lref\GSWI{
M.~B.~Green, J.~H.~Schwarz and E.~Witten, ``Superstring Theory. Vol. 1: Introduction,''
}

\lref\dineseiberg{
M.~Dine and N.~Seiberg, ``Is The Superstring Weakly Coupled?,'' Phys.\ Lett.\ B {\bf 162}, 299 (1985).
}

\lref\giddings{
S.~B.~Giddings, ``The fate of four dimensions,'' Phys.\ Rev.\ D {\bf 68}, 026006 (2003) [arXiv:hep-th/0303031].
}


\lref\dealwisetal{
R.~C.~Myers, ``New Dimensions For Old Strings,'' Phys.\ Lett.\ B {\bf 199}, 371 (1987).
S.~P.~de Alwis, J.~Polchinski and R.~Schimmrigk, ``Heterotic Strings With Tree Level Cosmological Constant,''
Phys.\ Lett.\ B {\bf 218}, 449 (1989).
}

\lref\APS{
A.~Adams, J.~Polchinski and E.~Silverstein, ``Don't panic! Closed string tachyons in ALE space-times,'' JHEP
{\bf 0110}, 029 (2001) [arXiv:hep-th/0108075].
 }

\lref\tachRGrefs{
C.~Vafa, ``Mirror symmetry and closed string tachyon condensation,'' arXiv:hep-th/0111051.
J.~A.~Harvey, D.~Kutasov, E.~J.~Martinec and G.~Moore, ``Localized tachyons and RG flows,''
arXiv:hep-th/0111154.
A.~Dabholkar and C.~Vafa, ``tt* geometry and closed string tachyon potential,'' JHEP {\bf 0202}, 008 (2002)
[arXiv:hep-th/0111155].
}

\lref\otherGRtach{
R.~Gregory and J.~A.~Harvey, ``Spacetime decay of cones at strong coupling,'' Class.\ Quant.\ Grav.\ {\bf 20},
L231 (2003) [arXiv:hep-th/0306146];
M.~Headrick, ``Decay of C/Z(n): Exact supergravity solutions,'' JHEP {\bf 0403}, 025 (2004)
[arXiv:hep-th/0312213].
 }

\lref\twistedcircles{
J.~R.~David, M.~Gutperle, M.~Headrick and S.~Minwalla, ``Closed string tachyon condensation on twisted
circles,'' JHEP {\bf 0202}, 041 (2002) [arXiv:hep-th/0111212].
 }

\lref\barton{
Y.~Okawa and B.~Zwiebach, ``Twisted tachyon condensation in closed string field theory,'' JHEP {\bf 0403}, 056
(2004) [arXiv:hep-th/0403051].
}

\lref\andytach{
M.~Gutperle and A.~Strominger, ``Fluxbranes in string theory,'' JHEP {\bf 0106}, 035 (2001)
[arXiv:hep-th/0104136].
 }

\lref\intriligatorlectures{
 K.~A.~Intriligator and N.~Seiberg, ``Lectures on supersymmetric gauge theories and electric-magnetic
duality,'' Nucl.\ Phys.\ Proc.\ Suppl.\  {\bf 45BC}, 1 (1996) [arXiv:hep-th/9509066].
}

\lref\asymmetricorb{
K.~S.~Narain, M.~H.~Sarmadi and C.~Vafa, ``Asymmetric Orbifolds,'' Nucl.\ Phys.\ B {\bf 288}, 551 (1987);
A.~Dabholkar and J.~A.~Harvey,
JHEP {\bf 9902}, 006 (1999) [arXiv:hep-th/9809122];
K.~Aoki, E.~D'Hoker and D.~H.~Phong, ``On the construction of asymmetric orbifold models,''
arXiv:hep-th/0402134.
 }

\lref\nongeom{
S.~Hellerman, J.~McGreevy and B.~Williams, ``Geometric constructions of nongeometric string theories,'' JHEP
{\bf 0401}, 024 (2004) [arXiv:hep-th/0208174].
A.~Flournoy, B.~Wecht and B.~Williams, ``Constructing nongeometric vacua in string theory,''
arXiv:hep-th/0404217.
 }

\lref\andydim{
A.~Strominger, ``The Inverse Dimensional Expansion In Quantum Gravity,'' Phys.\ Rev.\ D {\bf 24}, 3082 (1981).
 }

 \lref\strjunc{
O.~Aharony, J.~Sonnenschein and S.~Yankielowicz, ``Interactions of strings and D-branes from M theory,'' Nucl.\
Phys.\ B {\bf 474}, 309 (1996) [arXiv:hep-th/9603009];
J.H. Schwarz, ``Lectures on Superstring and M theory Dualities: Given at ICTP Spring School and at TASI Summer
School", {\it Nucl. Phys. Proc. Suppl.} {\bf 55B} (1997)1, hep-th/9607201;
 O.~Aharony and A.~Hanany,
``Branes, superpotentials and superconformal fixed points,'' Nucl.\ Phys.\ B {\bf 504}, 239 (1997)
[arXiv:hep-th/9704170];
M.~R.~Gaberdiel and B.~Zwiebach, ``Exceptional groups from open strings,'' Nucl.\ Phys.\ B {\bf 518}, 151 (1998)
[arXiv:hep-th/9709013];
O.~DeWolfe and B.~Zwiebach, ``String junctions for arbitrary Lie algebra representations,'' Nucl.\ Phys.\ B {\bf
541}, 509 (1999) [arXiv:hep-th/9804210];
O.~Aharony, A.~Hanany and B.~Kol, ``Webs of (p,q) 5-branes, five dimensional field theories and grid diagrams,''
JHEP {\bf 9801}, 002 (1998) [arXiv:hep-th/9710116];
K.~Dasgupta and S.~Mukhi, ``BPS nature of 3-string junctions,'' Phys.\ Lett.\ B {\bf 423}, 261 (1998)
[arXiv:hep-th/9711094];
A.~Sen, ``String network,'' JHEP {\bf 9803}, 005 (1998) [arXiv:hep-th/9711130];
S.~J.~Rey and J.~T.~Yee, ``BPS dynamics of triple (p,q) string junction,'' Nucl.\ Phys.\ B {\bf 526}, 229 (1998)
[arXiv:hep-th/9711202];
O.~Bergman and B.~Kol, ``String webs and 1/4 BPS monopoles,'' Nucl.\ Phys.\ B {\bf 536}, 149 (1998)
[arXiv:hep-th/9804160].
}

\lref\BP{
R.~Bousso and J.~Polchinski, ``Quantization of four-form fluxes and dynamical neutralization of the cosmological
constant,'' JHEP {\bf 0006}, 006 (2000) [arXiv:hep-th/0004134].
}

\lref\HorowitzNW{ G.~T.~Horowitz and J.~Polchinski, ``A correspondence principle for black holes and strings,''
Phys.\ Rev.\ D {\bf 55}, 6189 (1997) [arXiv:hep-th/9612146].
}

\lref\domainwalls{
S.~Gukov, C.~Vafa and E.~Witten, ``CFT's from Calabi-Yau four-folds,'' Nucl.\ Phys.\ B {\bf 584}, 69 (2000)
[Erratum-ibid.\ B {\bf 608}, 477 (2001)] [arXiv:hep-th/9906070];
S.~Kachru, X.~Liu, M.~B.~Schulz and S.~P.~Trivedi, ``Supersymmetry changing bubbles in string theory,'' JHEP
{\bf 0305}, 014 (2003) [arXiv:hep-th/0205108].
}

\lref\tometal{
T.~Banks, ``A critique of pure string theory: Heterodox opinions of diverse  dimensions,'' arXiv:hep-th/0306074;
T.~Banks and W.~Fischler, ``An holographic cosmology,'' arXiv:hep-th/0111142.
}

\lref\bfss{
T.~Banks, W.~Fischler, S.~H.~Shenker and L.~Susskind, ``M theory as a matrix model: A conjecture,'' Phys.\ Rev.\
D {\bf 55}, 5112 (1997) [arXiv:hep-th/9610043].
}

\lref\BL{
D.~Berenstein and R.~G.~Leigh, ``String junctions and bound states of intersecting branes,'' Phys.\ Rev.\ D {\bf
60}, 026005 (1999) [arXiv:hep-th/9812142].
}

\lref\SW{
L.~Susskind and E.~Witten, ``The holographic bound in anti-de Sitter space,'' arXiv:hep-th/9805114.
}

\lref\AcharyaII{ B.~S.~Acharya, F.~Denef, C.~Hofman and N.~Lambert, ``Freund-Rubin revisited,''
arXiv:hep-th/0308046.
}

\lref\moore{G. Moore, in progress}

\lref\alexetal{
A.~Saltman and E.~Silverstein, ``The scaling of the no-scale potential and de Sitter model building,''
arXiv:hep-th/0402135.
 }

\lref\Douglas{
S.~Ashok and M.~R.~Douglas, ``Counting flux vacua,'' arXiv:hep-th/0307049.
}

\lref\SKunp{S. Kachru, unpublished}

\lref\kklt{
S.~Kachru, R.~Kallosh, A.~Linde and S.~P.~Trivedi, ``De Sitter vacua in string theory,'' arXiv:hep-th/0301240.
}

\lref\MSS{
A.~Maloney, E.~Silverstein and A.~Strominger, ``De Sitter space in noncritical string theory,''
arXiv:hep-th/0205316;
E.~Silverstein, ``(A)dS backgrounds from asymmetric orientifolds,'' arXiv:hep-th/0106209.
}

\lref\achar{
B.~S.~Acharya, ``A moduli fixing mechanism in M theory,'' arXiv:hep-th/0212294.
}

\lref\GKP{
S.~B.~Giddings, S.~Kachru and J.~Polchinski, ``Hierarchies from fluxes in string compactifications,'' Phys.\
Rev.\ D {\bf 66}, 106006 (2002) [arXiv:hep-th/0105097].
}

\lref\W{
E.~Witten, ``Anti-de Sitter space and holography,'' Adv.\ Theor.\ Math.\ Phys.\  {\bf 2}, 253 (1998)
[arXiv:hep-th/9802150].
}

\lref\StromingerSH{ A.~Strominger and C.~Vafa, ``Microscopic Origin of the Bekenstein-Hawking Entropy,'' Phys.\
Lett.\ B {\bf 379}, 99 (1996) [arXiv:hep-th/9601029].
}

\lref\GKPads{
S.~S.~Gubser, I.~R.~Klebanov and A.~M.~Polyakov, ``Gauge theory correlators from non-critical string theory,''
Phys.\ Lett.\ B {\bf 428}, 105 (1998) [arXiv:hep-th/9802109].
}

\lref\KachruNS{ S.~Kachru, X.~Liu, M.~B.~Schulz and S.~P.~Trivedi, ``Supersymmetry changing bubbles in string
theory,'' JHEP {\bf 0305}, 014 (2003) [arXiv:hep-th/0205108].
}

\lref\GH{
G.~W.~Gibbons and S.~W.~Hawking, ``Cosmological Event Horizons, Thermodynamics, And Particle Creation,'' Phys.\
Rev.\ D {\bf 15}, 2738 (1977).
}

\lref\DS{
M.~Fabinger and E.~Silverstein, ``D-Sitter space: Causal structure, thermodynamics, and entropy,''
arXiv:hep-th/0304220.
}

\lref\juan{
J.~M.~Maldacena, ``The large N limit of superconformal field theories and supergravity,'' Adv.\ Theor.\ Math.\
Phys.\ {\bf 2}, 231 (1998) [Int.\ J.\ Theor.\ Phys.\  {\bf 38}, 1113 (1999)] [arXiv:hep-th/9711200].
}

\lref\dSCFT{
A.~Strominger, ``The dS/CFT correspondence,'' JHEP {\bf 0110}, 034 (2001) [arXiv:hep-th/0106113]
}

\lref\dSObj{
E.~Witten, ``Quantum gravity in de Sitter space,'' arXiv:hep-th/0106109;
W.~Fischler, A.~Kashani-Poor, R.~McNees and S.~Paban, ``The acceleration of the universe, a challenge for string
theory,'' JHEP {\bf 0107}, 003 (2001) [arXiv:hep-th/0104181];
S.~Hellerman, N.~Kaloper and L.~Susskind, ``String theory and quintessence,'' JHEP {\bf 0106}, 003 (2001)
[arXiv:hep-th/0104180];
A.~Strominger, ``The dS/CFT correspondence,'' JHEP {\bf 0110}, 034 (2001) [arXiv:hep-th/0106113].
N.~Goheer, M.~Kleban and L.~Susskind, ``The trouble with de Sitter space,'' JHEP {\bf 0307}, 056 (2003)
[arXiv:hep-th/0212209].
L.~Dyson, J.~Lindesay and L.~Susskind, ``Is there really a de Sitter/CFT duality,'' JHEP {\bf 0208}, 045 (2002)
[arXiv:hep-th/0202163];
T.~Banks, W.~Fischler and S.~Paban, ``Recurrent nightmares?: Measurement theory in de Sitter space,'' JHEP {\bf
0212}, 062 (2002) [arXiv:hep-th/0210160];
T.~Banks and W.~Fischler, ``M-theory observables for cosmological space-times,'' arXiv:hep-th/0102077;
L.~Susskind, ``The anthropic landscape of string theory,'' arXiv:hep-th/0302219.

}

\lref\stringstalk{E. Silverstein, talk at Strings 2003}


\lref\KLT{
P.~Kraus, F.~Larsen and S.~P.~Trivedi, ``The Coulomb branch of gauge theory from rotating branes,'' JHEP {\bf
9903}, 003 (1999) [arXiv:hep-th/9811120].
}

\lref\AbbottQF{ L.~F.~Abbott, ``A Mechanism For Reducing The Value Of The Cosmological Constant,'' Phys.\ Lett.\
B {\bf 150}, 427 (1985).
}

\lref\BanksMB{ T.~Banks, M.~Dine and N.~Seiberg, ``Irrational axions as a solution of the strong CP problem in
an eternal universe,'' Phys.\ Lett.\ B {\bf 273}, 105 (1991) [arXiv:hep-th/9109040].
}

\lref\BrownKG{ J.~D.~Brown and C.~Teitelboim, ``Neutralization Of The Cosmological Constant By Membrane
Creation,'' Nucl.\ Phys.\ B {\bf 297}, 787 (1988).
}

\lref\FengIF{ J.~L.~Feng, J.~March-Russell, S.~Sethi and F.~Wilczek, ``Saltatory relaxation of the cosmological
constant,'' Nucl.\ Phys.\ B {\bf 602}, 307 (2001) [arXiv:hep-th/0005276].
}

\input epsf
\noblackbox
\newcount\figno
\figno=0
\def\fig#1#2#3{
\par\begingroup\parindent=0pt\leftskip=1cm\rightskip=1cm\parindent=0pt
\baselineskip=11pt \global\advance\figno by 1 \midinsert \epsfxsize=#3  \centerline{\epsfbox{#2}} \vskip 12pt
{\bf Fig.\ \the\figno: } #1\par
\endinsert\endgroup\par
}
\def\figlabel#1{\xdef#1{\the\figno}}


\Title{\vbox{\baselineskip12pt\hbox{hep-th/0405068} \hbox{SLAC-PUB-10441}\hbox{SU-ITP-04/18} }}{\vbox{\vskip
-1.5cm\centerline {TASI/PiTP/ISS Lectures} \vskip 0.5cm \centerline{On Moduli and Microphysics }}}

\vskip -0.3cm

\bigskip

\centerline{Eva Silverstein\footnote{$^*$} {SLAC and Department of Physics, Stanford University, Stanford, CA
94309}}

\vskip .3in \centerline{\bf Abstract} { } \vskip .3in

I review basic forces on moduli that lead to their stabilization, for example in the supercritical and KKLT
models of de Sitter space in string theory, as well as an $AdS_4\times S^3\times S^3$ model I include which is
not published elsewhere. These forces come from the classical dilaton tadpole in generic dimensionality,
internal curvature, fluxes, and branes and orientifolds as well as non-perturbative effects. The resulting (A)dS
solutions of string theory make detailed predictions for microphysical entropy, whose leading behavior we
exhibit on the Coulomb branch of the system. Finally, I briefly review recent developments concerning the role
of velocity-dependent effects in the dynamics of moduli.  These lecture notes are based on material presented at
various stages in the 1999 TASI, 2002 PiTP, 2003 TASI, and 2003 ISS schools.

\smallskip

\newsec{Introduction and Motivation}

The large scale dynamics of the universe (including the question of whether the universe is large at all in a
given model) is determined to a significant extent by the cosmological term in the effective action.  In a
theory such as string theory with scalar fields, this term is in general a nontrivial function of the scalar
field VEVs, which in turn determine the coupling constants appearing in low energy particle physics in each
vacuum. Acceleration of the universe is obtained if the scalar fields roll slowly enough that their kinetic
energy is dominated by their potential energy.\foot{which is not precisely the same condition as the ``slow roll
conditions".} This situation can arise if the moduli are stabilized or metastabilized, if the potential is very
flat, or if significant interactions slow down the scalar motion.

The observed geometry of the universe suggests two phases of accelerated expansion--one in the past (inflation)
and one beginning to dominate now. Observations also strongly bound time evolution of fundamental constants, and
extra scalar gravitational-strength forces. It is therefore important to determine the contributions to the
cosmological term and understand how the scalar fields become stabilized. It has recently become clear that
other effects dependent on the scalar field velocities can play a large role in their dynamics, and it is
important to incorporate these kinetic effects as well.  In this set of notes, I will review some of the recent
progress concerning the stabilization of moduli and the resulting physics of dark energy and inflation.  While
there has been much progress, there are also larger open problems--both conceptual and technical--making this a
valuable subject to get into at this stage.

Although the cosmological term is dominant at long distances, it is sensitive to microphysics in various ways.
Firstly, the quantum corrections to the cosmological term are UV sensitive, coming at the scale of SUSY breaking
(which may be anywhere from the TeV scale up to the string or Planck scale).  Secondly, the forces available to
fix the moduli depend on the possible ingredients in the theory; for example string theory in a given background
determines the spectrum of low energy fields and in principle determines the zoo of branes and other defects
which are consistent physical objects within the theory.  Finally, as in black hole physics, there is some
evidence that solutions with positive cosmological constant have an associated horizon entropy and
temperature.\foot{Here we will find metastable solutions with positive cosmological constant, so these
thermodynamic parameters are at best approximate; however one finds that equilibrium can be attained in a much
shorter time than the decay time out of the metastable vacuum.} If it is accounted for by microphysical degrees
of freedom, this may provide a dual description of the dark energy, or some aspects of it, in terms of the
degrees of freedom of string theory.

For all these reasons, it is important to get a handle on moduli stabilization and microphysics.  In what
follows, I will overview some basic aspects of this subject.  I will not reproduce all of the calculations I
covered in the lectures, leaving out some of the ones readily available in existing papers but including here a
set of simple but important computations required for getting started in this area. I will otherwise try to
provide an overview of the works I covered in the lectures, emphasizing some points that are especially
important and sometimes confusing to those entering this area of research.\foot{It is also practically
impossible to include all the important references here, but I hope to have included a representative sample of
the recent (as well as some older) efforts in this direction.  Earlier TASI lecture notes on moduli and
cosmology, such as \othermodnotes\ are also available.}

\newsec{Basic Contributions}

In this section we will catalogue a number of contributions to the moduli potential.  In a generic situation,
these contributions will all be involved (plus others not covered here, depending on the circumstances). As we
will see, there are more than enough contributions to provide balancing forces on all the moduli, and we expect
the moduli to be fixed in many cases, leading to dS or AdS solutions.

Before delving into the detailed contributions, let us start by noting some important generalities concerning
moduli stabilization, including the rescaling to Einstein frame and the behavior of the moduli potential near
large radius and weak coupling limits. Then we will list a bunch of important contributions to the $4d$
potential energy, and note how they scale with moduli. We will discuss the issue of tachyons, and briefly
introduce the method of fixing moduli by orbifolding.  In \S3, we will put various subsets of these ingredients
together to provide compensating forces on the moduli.

\subsec{General Considerations}

One basic but important aspect to keep track of is the ``frame".  It is often most convenient to rescale the
metric so that the Einstein term $\int \sqrt{g}{\cal R}$ in the action has no moduli dependence in its
coefficient. This is not what arises immediately from dimensional reduction and perturbative string
calculations. For example, the worldsheet path integral leads to tree level amplitudes in closed string theory,
such as those determining the Einstein-Hilbert term in the Lagrangian, which arise at order $1/g_s^2$ in terms
of the string coupling $g_s$ (i.e. coming with a power of $e^{-2\Phi}$ where $\Phi$ is the dilaton). In the
presence of moduli dependence in the Einstein term--of the general form $\int f(\phi_I){\cal R}^{(s)}$--the
problem of solving the equations of motion for the scalars $\phi_I$ has an extra tadpole from the spacetime
curvature.

Instead, as we will derive presently, we can remove this dependence by working with a rescaled metric to obtain
the canonical Einstein term $\int \sqrt{g_E}{\cal R}_E$ where the subscript $E$ refers to the Einstein frame.
Then the scalar field equation will be simply $\del_\mu(\sqrt{g_E}g_E^{\mu\nu}\del_\nu\phi^I)=-\del_I {\cal
V}_E(\phi)=0$ where ${\cal V}_E$ is the Einstein frame potential energy.

Start in the $D$-dimensional string frame, for some $D>4$ which we plan to compactify down to four dimensions
(or more generally down to some other dimension $d$ in toy models).  The action in string frame is
\eqn\DdimR{{\cal S}=\int d^Dx\sqrt{g^{(10)}_s}{e^{-2\Phi}\over l_s^8}{\cal R}_s+\dots}
where $l_s$ is the string length, i.e. the square root of the inverse string tension. Reducing this to four
dimensions, on a $D-4$ dimensional space $X$ of volume $l_s^{D-4}V_X$, we obtain
\eqn\Sfourd{{\cal S}=\int d^4x\sqrt{g_s^{(4)}}{{e^{-2\Phi}V_X}\over l_s^2}{\cal R}_s+\dots}
Now let us write $e^\Phi=g_se^{\tilde\Phi}$ and $V_X=V_{X0}\tilde V_X$, with $g_s,V_{X0}$ constant and with the
four dimensional Planck length defined by
\eqn\Planck{l_4^2=l_s^2{g_s^2\over V_{X0}^2}}
This leaves $\tilde\Phi$ and $\tilde V_X$ as fluctuating scalars in four dimensions, the dependence on which we
wish to remove from the Einstein term.  We can do this by rescaling the four dimensional metric via
\eqn\metricrescaling{g_{\mu\nu,s}^{(4)}\equiv {e^{2\tilde\phi}\over \tilde V_X}g_{\mu\nu,E}^{(4)}}
which defines the Einstein frame metric $g_{\mu\nu,E}^{(4)}$. With this rescaling, our action has become
\eqn\EframeE{{\cal S}=\int d^4x\sqrt{g_E^{(4)}}{1\over l_4^2}{\cal R}_E}
with the dimensionful coupling $l_4^2=G_N$ appearing as it should but with all scalar field dependence removed.
This rescaling similarly removes the $e^{-2\Phi}$ dependence in front of the $\Phi$ kinetic term, which leaves
us with an exponential potential in terms of the canonically normalized field, as we will see next.

As I alluded to above, this rescaling affects also the form of the potential energy, in an important way. Namely
the string-frame action dimensionally reduced to four dimensions also contains potential energy terms, which
will be one of our main concerns in these notes.  If we write the potential energy term in the action as
\eqn\potS{\int d^4x\sqrt{g_s^{(4)}}{1\over l_s^4}{\cal V}_s^{(4)} }
(i.e. defining ${\cal V}_s^{(4)}$ to be the string frame potential energy in string units), it becomes in terms
of the Einstein frame metric
\eqn\potE{\int d^4x\sqrt{g_E^{(4)}} {1\over l_4^4}{e^{4\Phi}\over V_X^2}{\cal V}_s^{(4)}}
Here we used the rescaling \metricrescaling\ in the $\sqrt{g}$ factor in \potS, and also wrote the expression in
terms of the $4d$ Planck length $l_4$ rather than the string length $l_s$.

So finally we can identify the four dimensional Einstein frame cosmological term to be
\eqn\ELam{\Lambda_E^{(4)}={1\over l_4^2}{e^{4\Phi}\over V_X^2}{\cal V}_s^{(4)}}

Similar comments would apply to any situation with a function $f$ of moduli multiplying the curvature; one can
transform to Einstein frame via the relations \metricrescaling\ELam\ with the replacement of $V_X/e^{2\Phi}$ by
$f$.

The resulting volume and coupling dependence in the potential in Einstein frame leads to an important feature of
the physics: all sources of energy coming from local physics on the compact space scale to zero as the coupling
approaches zero and/or the volume approaches infinity (as discussed in more detail in e.g.
\dineseiberg\giddings).  Even if there is an energy density filling the full $D$-dimensional spacetime, the two
powers of $V_X$ arising in the denominator from the conversion to Einstein frame beat the power of $V_X$ from
integrating the $D$-dimensional energy density, producing a contribution that approaches zero as $V_X\to\infty$.

As a result, in order to stabilize the moduli expanding about a controlled weakly coupled and/or large radius
solution, one requires both negative and positive sources of potential energy, with sufficiently large
coefficients in front of the subleading terms in the expansion to produce a balance of forces within the weakly
coupled regime.

In addition to these runaway moduli, there are in general also periodic moduli, coming from integrated
$p-1$-form potentials.  Although they cannot run away to infinity and destabilize a compactification, the
potential for the runaway moduli generally depends on them.

Solving the scalar equation of motion at a local minimum $V(\phi_*)$ yields a cosmological constant
$\Lambda_*=V_*/M_P^2$. The classical solution for the metric is then de Sitter space for $V_*>0$, Minkowski
space for $V_*=0$, and anti-de Sitter space for $V_*<0$. The global structures of these three spaces are quite
different, so a small positive value is qualitatively distinct from a precisely zero energy vacuum or a small
negative one.

The simplest two possibilities are indicated in figures 1-2.  In figure 1, one obtains an AdS solution in
expansion about weak coupling by playing a negative contribution off of a positive contribution at higher order.
In figure 2, one obtains a dS solution by playing a leading order positive contribution off of a subleading
negative and a further subleading positive contribution.  In all cases, a necessary condition for control is
that the coefficients of the subleading terms be $>>1$ so that the coupling is stabilized at a weak value.

\global\advance\figno by 1

\ifig\tasiI{This figure illustrates the simplest way to stabilize the moduli near weak coupling with a negative
cosmological constant.  The horizontal axis is the coupling and the vertical is the potential.}
{\epsfxsize4in\epsfbox{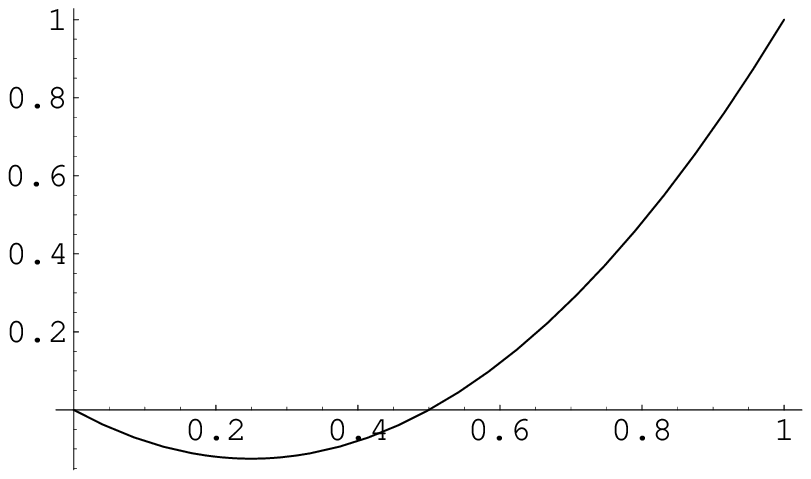}}

\ifig\tasiII{This figure illustrates the simplest way to stabilize the moduli at positive cosmological constant
near weak coupling.  At least three independent terms in an expansion about weak coupling are required, with the
middle term negative.} {\epsfxsize4in\epsfbox{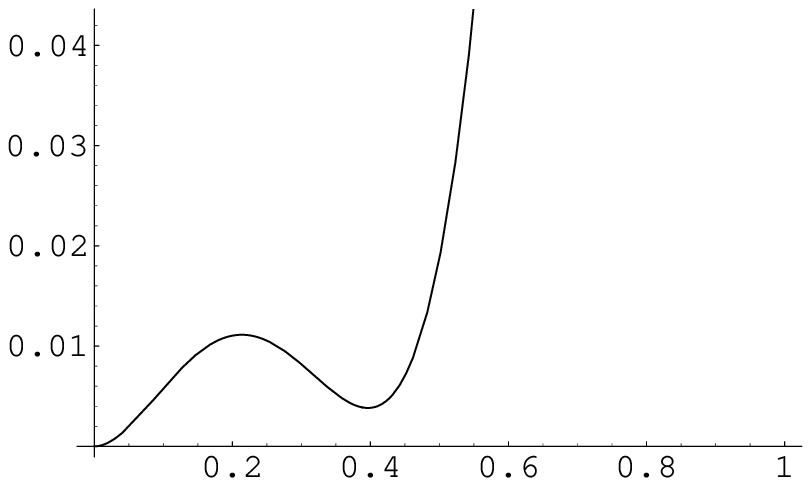}}

By using the known ingredients in string theory, we will find that with ${\cal N}\le 1$ supersymmetry in four
dimensions we can find compensating forces on all the moduli, and strong evidence that they can be arranged to
yield many vacua with a finely spaced ``discretuum" of possible values for $V_*$. Genericity arguments along the
lines of \BP\ suggest the presence of small nonzero cosmological constant vacua, but don't guarantee (or even
suggest) the presence of vacua with broken supersymmetry and precisely zero cosmological constant in four
dimensions.

Another general feature of the moduli potential is that it is extremized at enhanced symmetry points. These are
special loci in $\phi$ space where an extra symmetry appears.  Expanding around such a point $\vec
\phi=\vec\phi_*+\delta\vec\phi$, where $\vec\phi_*$ is the enhanced symmetry point (ESP) and $\delta\vec\phi$
transforms nontrivially under the symmetry, one cannot have a term in the potential linear in $\delta\vec\phi$.
Hence the ESP is at least an extremum of the effective potential in the directions $\delta\vec\phi$ acted on by
the symmetry. Some of these extrema are local minima, and some are saddle points or local maxima of the
effective potential. It was suggested in e.g. \dinesymm\ that such points may play a special role in string
cosmology. Recently, dynamical mechanism trapping kinetic-dominated scalar fields at such ESPs has emerged
\trapping, as have results suggesting a distribution of solutions clustered near certain ESP loci
\clusteringvac.

\subsec{Dimensionality and the tree level cosmological term}

Let us begin our enumeration of contributions to the cosmological term in string theory with perhaps the most
basic contribution.  Perturbative string theory can be formulated in any dimensionality, with the ``critical
dimension" having the property that the classical theory has exactly flat solutions in the prescribed dimension
($D_{crit}=10$ for the superstring, $D_{crit}=26$ for the bosonic string).

Since exactly flat spacetime is non-generic and unrealistic, this alone does not argue for the critical
dimension.  It is possible however, and widely assumed, that supersymmetry provides a reason to specialize to
this case. In particular, in the case of string theories with worldsheet supersymmetry, the flat space solution
of the critical theory also has extended supersymmetry in spacetime, which reduces on appropriate
compactifications to theories with supersymmetric low energy effective actions governing the dynamics in four
dimensions.  This symmetry provides an elegant solution to the problem of understanding the hierarchy between
the Planck and electroweak scales, contributes a natural dark matter candidate, and facilitates the unification
of forces into grand unified theories--and therefore is a leading contender for particle physics beyond the
Standard Model. However, at this writing it remains an open experimental problem to determine whether
supersymmetry plays this role in nature; no direct evidence exists to decide this question. Independently of
that, supersymmetry simplifies the analysis of quantum field theory and string theory.  It is not known whether
or not the noncritical strings have supersymmetric vacua, though as we will discuss below there are some reasons
to expect that at least the critical superstring theories arise as vacua of some of the supercritical string
theories.

In any case, in order to understand what (if anything) the theory as a whole might predict, and more modestly to
understand the full range of known model building possibilities, it is important to consider the string theory
vacua in generic dimensions. In $D$ dimensions, one finds a contribution to the tree level dilaton potential of
order
\eqn\Dimterm{\Delta {\cal V}_s^{(4)}\propto {(D-D_{crit})\over l_s^4} e^{-2\Phi}V_X}
as discussed in \dealwisetal\joeI\GSWI.  In four-dimensional Einstein frame, following our derivation \ELam\
this is
\eqn\DimtermE{\Delta\Lambda_E\propto {(D-D_{crit})\over l_4^2} {e^{2\Phi}\over {V_X}}}
This follows from studying the worldsheet reparameterization and Weyl symmetries in the presence of an arbitrary
number of free matter scalars $X^M$, $M=0,\dots,D-1$; one finds a contribution to the trace of the worldsheet
stress-energy tensor proportional to $D-D_{crit}$. Following the discussion in \GSWI, starting from the flat
worldsheet metric $\eta_{\alpha\beta}$ we could fix classically, we vary it to
$g_{\alpha\beta}=\eta_{\alpha\beta}+h_{\alpha\beta}$, whereupon the worldsheet action changes by
\eqn\shiftworldsheet{\delta S_{ws}=\int d^2z T_{\alpha\beta}h^{\alpha\beta}}
where $T_{\alpha\beta}$ is the worldsheet stress-energy tensor. After this deformation, the stress energy tensor
shifts, in particular
\eqn\shiftT{\delta<T_{zz}>=<T_{zz}\int h_{\alpha\beta}T^{\alpha\beta}>\sim (D-D_{crit})\int {h_{\bar z\bar
z}\over z^4}}
to leading order in the deformation. Now if at this order $T_{\bar z z}$ were still zero, we would lose
energy-momentum conservation since
\eqn\Tcons{\del^\alpha T_{\alpha\beta}=\del^z T_{zz}+\del^{\bar z}T_{\bar z z}}
Since the first term here is nonzero, so must be the second term to preserve energy-momentum conservation on the
string worldsheet.  This leads to a covariant contribution
\eqn\trace{T^\alpha_\alpha=-{(D-D_{crit})\over 12}{\cal R}^{(2)}}
where ${\cal R}^{(2)}$ is the worldsheet curvature scalar.  In other words, in a flat spacetime background with
constant dilaton, the theory would not preserve Weyl invariance.  However, in general there are other
contributions, such as that generated by the worldsheet coupling $\int \Phi {\cal R}^{(2)}$.  Taking this into
account, one finds a dilaton equation of motion
\eqn\fulltrace{T^\alpha_\alpha=-{1\over 2}{\cal R}^{(2)}({(D-D_{crit})\over 6}-\alpha^\prime
\nabla^2\Phi+\alpha^\prime(\nabla\Phi)^2+\dots)}
where the $+\dots$ refers to higher orders in the $\alpha^\prime$ and string loop expansion and contributions
from other fields such as curvatures and fluxes (to be discussed below). The terms in \fulltrace\ arise from a
cosmological term of the form \DimtermE\ in the low energy effective action.

One would like to know how these noncritical theories fit into the web of M theory backgrounds.  It is likely
that they are related by closed string tachyon condensation, related to turning on relevant perturbations on the
string worldsheet (see the recent paper \simeon\ for an interesting example of this in some concrete heterotic
models). As discussed in \dealwisetal, worldsheet renormalization group flows are related to spacetime
instabilities; condensing tachyons leads to reduction in the worldsheet degrees of freedom in the matter sector.
We see that in the above result: condensing a tachyon would reduce the potential energy, which correlates with
reducing $D-D_{crit}$.  In the next subsection, we will see that positive spacetime curvature has the same
effect, so that compactifying the supercritical theories on positively curved spaces such as spheres may reduce
them to critical theories.\foot{J. Polchinski has developed this relation more precisely in some cases.}

\subsec{Spatial Curvature in the Compactification}

Starting from the Einstein term in $D$ dimensions, dimensionally reducing over $X$ produces a contribution to
the four-dimensional effective action which scales like
\eqn\curvs{\Delta\Lambda_s\sim -\int d^4x\sqrt{g_s^{(4)}}\int_X {\cal R}^{(X)}{e^{-2\Phi}\over l_s^8}}
in string frame, and correspondingly rescaled as in \ELam\ in Einstein frame.  The sign here comes from the fact
that the gravitational action \DdimR\ contributes to the {\it potential} energy term with a minus sign.  For
example, if we compactify on an $S^6$ of volume $V_Xl_s^6=L^6l_s^6$ we obtain
\eqn\REframe{\Delta\Lambda_E\sim -{1\over l_4^2}{1\over L^2}{e^{2\Phi}\over V_X}}
in four dimensional Einstein frame, using the fact that the curvature scales like $1/L^2$ and positive curvature
contributes negatively to the potential energy term in the Lagrangian.  (Note that the conversion to Einstein
frame does not act on the internal metric or curvature.)  In general, it is worth remarking on the sign:
negative curvature contributes positively to the four dimensional potential energy, while positive curvature
contributes negatively.

\subsec{Fluxes}

An important contribution to the moduli potential comes from fluxes on cycles of the compactification (see
\oldflux\GVW\gkpetc\MSS. for some of the original papers on this method for lifting the moduli in string
theory). These are quantized: in terms of a p-form field strength $F_p$ on a p-cycle $C_p$ we get
\eqn\fluxquant{{1\over{2\pi l_s^{p-1}}}\int_{C_p}F_p=Q_p}
with integer $Q_p$.  This contributes positive potential energy via the flux kinetic terms proportional to
$F_p^2\sim Q^2/(V_p^2l_s^2)$, where $V_p$ is the volume of the p cycle in string units (i.e. the volume divided
by $l_s^p$).  Here we are taking a uniform flux on the $p$-cycle. For NS fields this kinetic term arises at the
closed string tree level, multiplied by $e^{-2\Phi}$ in string frame. RR fluxes have no $e^{\Phi}$ factors in
front of their kinetic terms in string frame.

We can now calculate the contribution to $\Lambda$ of the flux kinetic terms in Einstein frame taking into
account the results of \S2.1.  For example, for NS 3-form flux $H$ or RR 3-form flux $F$ on a compactification
from 7 to 4 dimensions on a 3-sphere of volume $V_3l_s^3$ one finds respectively
\eqn\fluxpotH{\Delta \Lambda_E\sim {{e^{2\Phi}Q_H^2}\over{l_4^2V_3^3}}}
and
\eqn\fluxpotF{\Delta \Lambda_E\sim {{e^{4\Phi}Q_F^2}\over{l_4^2V_3^3}}}
Note that the fluxes naturally force the cycles to expand, lowering the flux energy density on the cycles. Note
also that the RR fluxes come at the same order as 1-loop effects.

There are many choices of fluxes on a typical compactification, leading to huge numbers of vacua as discussed
first in \BP.  We will explain a simple estimate for this number in flux compactifications when we get to the
models in \S3.

There are important couplings involving $F_3$ and $H_3$ which I should note.  Firstly, we must take into account
axion couplings.  In the type IIB Lagrangian in 10 dimensions, we have a term proportional to
\eqn\FH{\int d^{10}x {\sqrt{g_s^{(10)}}\over{l_s^8}}|F-C_0H|^2}
where $C_0$ is the IIB axion \joeII.  So although the axion is a periodic variable, not subject to runaway to
infinity, its VEV affects the forces on the other moduli including the potential runaway directions such as the
dilaton. Also, there is an important topological restriction on compact manifolds coming from a contribution of
3-form fluxes to 3-brane charge: the Gauss' law relation between $L\sim \int H\wedge F$ and orientifold 3-plane
and D3-brane charge is
\eqn\gauss{{1\over{2(2\pi)^4(\alpha')^2}}\int H\wedge F = {1\over
4}(N_{O3}-N_{\overline{O3}})-N_{D3}+N_{\overline{D3}}}
This contribution to the 3-brane charge from $\int H\wedge F$ comes from a coupling $\int C_4\wedge F\wedge H$
between the 3-brane gauge potential $C_4$ and $H\wedge F$. In the next section we will study the contribution of
branes to the moduli potential.

\subsec{Orientifolds and D-branes}

Next let us turn to the effects of defects in the compactification.  In particular, Dp-branes and orientifolds
contribute to the low energy effective potential via their tension.  From their string frame tension
\joeI\joeII, $T_{Dp,s}\sim {1\over {g_sl_s^{p+1}}}$, a Dp-brane at a point in a compactification on $X$ with
volume $V_X$ in string units produces a 4d Einstein frame potential energy
\eqn\DbraneE{\Delta \Lambda_E\sim {e^{3\Phi}\over{l_4^2V_X^2}}}
Similarly, orientifold planes make a contribution scaling the same parameterically as a function of the coupling
and volume, but with an opposite sign.\foot{In general dimensions, we can study how the tension of orientifold
planes depends on the dimensionality $D$ as well.  This was done in \MSS, where we found that in a toroidal
orientifold, the total contribution from the order $2^D$ orientifold fixed planes scales like $2^{D/4}$.} This
negative contribution from orientifolds will be crucial, and makes sense physically because orientifolds involve
a $Z_2$ identification which changes the asymptotic geometry far from the object--hence they cannot be
dynamically produced. Relatedly, their number in any given compactification is bounded.  This places interesting
limitations on stabilization of moduli, since the strength of the negative contribution to the potential in
figure 2 is important for determining how small the cosmological constant can be.

As observed and used in \kklt, in the presence of warping in a compactification, defects localized in regions of
small warp factor have an extra suppression to their tension.  This provides a natural way in which these
objects can provide forces competing with contributions which are a priori more subleading, such as perturbative
and non-perturbative quantum corrections.

\subsec{Loop Effects}

Although they appeared at different orders in $g_s\sim e^\Phi$, the contributions we discussed so far are all
classical energy densities.  Quantum corrections also generically arise and can compete with the other effects.

At one loop in perturbation theory, in the presence of ${\cal N}=1$ or ${\cal N}=0$ supersymmetry, the moduli
potential receives corrections.  The computation of the 1-loop potential is discussed in detail in \joeI\
(chapter 7), whose description I largely worked through in the lectures at the schools but will not reproduce
here. The result is UV sensitive, generalizing the field theory result
 $\Lambda_{1-loop}=\int{d^3\vec k\over (2\pi)^3}\omega_{\vec k}$ where $\omega_{\vec k}$ is the frequency of
 the mode solutions.  As discussed above, in flux models the RR fluxes arise at the same order as this 1-loop term,
 and can be used to tune the total contribution close to a desired value.

\subsec{Non-Perturbative Effects}

In addition to the perturbative contributions, one has non-perturbative corrections coming from instantons,
gaugino condensation, and other effects.  At the level of a superpotential in models with a supersymmetric low
energy effective field theory, such effects can be the leading ones lifting some of the moduli.  Many string
theories reduce at low energies to the quantum field theories containing the field content producing instantons
or gaugino condensation (with the obvious caveat that non-perturbative effects from strong coupling gauge
dynamics occurs only in cases with asymptotically free matter content). There have been several good lecture
series on this subject, which I will not cover in any detail here (see e.g. \intriligatorlectures).  The moduli
dependence of such effects comes in part from the gauge theory relation $\Lambda_{QCD}\sim e^{-C/g_{YM}^2}$
where $C$ is an order 1 constant and $g_{YM}^2$ can depend on volumes of cycles as well as the string coupling.
For example, 7-branes wrapped on 4-cycles of linear size $Ll_s$ have $g_{YM}^2\sim g_s/L^4$.

Nonperturbative effects play an important role in the KKLT construction, to be overviewed in the next section.
Specific examples in this context were recently studied in \instegs.

Also at the classical level, one can have worldsheet instantons which produce corrections non-perturbative in
$\alpha^\prime$ \dsww.

\subsec{Tachyons}

In addition to tadpole forces on the moduli, one often encounters a situation with a scalar field $T$ in the
string theory spectrum whose mass squared is negative.  This indicates an instability.  In quantum field theory,
in this situation one would calmly shift the VEV of $T$ until it reached a metastable vacuum, if one exists in
the theory; otherwise one would discard the theory. In perturbative string theory, at the string scale, one is
usually stuck with a first quantized description of the unstable maximum, and we generally lack a sufficiently
global view of the space of string vacua to determine immediately if $T$ may be shifted away from the unstable
maximum at $T=0$ to a metastable position.

The worldsheet point of view on tachyon condensation is instructive.  An on-shell tachyon state is a solution to
the worldsheet constraint $L_0 |\Psi>=0$ producing the mass shell condition with $k^2=m_T^2<0$. Turning it on at
zero momentum $k=0$ would correspond to a relevant perturbation on the string worldsheet, which results in a
decrease in the number of degrees of freedom.

In recent years progress was made on this question for tachyons associated to D-branes (covered in Sen's
lectures), and localized tachyons in the twisted closed string spectrum of orbifold theories.  For example \APS,
consider orbifolds of the type II string in which we make a $Z_k$ identification with generator
\eqn\orbaction{g: ~~~ z_j\to e^{2\pi n_j i/k}z_j}
on complex coordinates $z_j$, $j=1,\dots,d/2$ for some integers $n_j$.  This produces a cone, with tip at
$z_j=0$ and a base $S^{d-1}/Z_k$ determined by the action \orbaction\ on the angular directions in $\IC^{d/2}$.
The spectrum of closed strings includes those twisted by $g^n,n=1,\dots,k-1$.  That is, in the $nth$ twisted
sector, the worldsheet scalars $Z_j(\sigma,\tau)$ corresponding to the spacetime coordinates $z_j$ satisfy
\eqn\twistedsect{Z_j(\sigma+2\pi,\tau)=g^nZ_j(\sigma)}
with the spacetime fermions satisfying corresponding relations.  This twisting of the boundary conditions of the
closed string leads to a modification of the ground state energy in the twisted sectors, and generically one
finds tachyons in the twisted sectors.

The tip of the cone is a highly curved region, for which low energy field theory and general relativity are not
sufficient. This is one of the simplest examples of a naked singularity which is resolved by string theory, at
least in the cases where the deficit angles in the cone are orbifold angles. In these situations, D-brane probes
can be used to determine the leading effect of condensing $T$; as described in \APS, this leads to decay into
$Z_{k^\prime}$ orbifolds for $k^\prime<k$ and matches to solutions in which a shell of graviton-dilaton solution
interpolates between the $Z_{k^\prime}$ conical tip and the $\IR\times S^{d/2}/\IZ_k$ asymptotics (as studied in
\APS\otherGRtach). Other techniques use the power of worldsheet supersymmetry to control the RG flow for the
lowest tachyons in these models \tachRGrefs.  In addition, one can regulate the situation with an extra shift,
producing examples amenable to a low energy field theory analysis \twistedcircles.  Finally string field theory
has been successfully applied to this process in \barton.

It is still an open question to determine the fate of bulk (as opposed to localized) tachyons.  Some interesting
ideas were proposed recently in \andytach\twistedcircles\IgorRR.  The proposal in \IgorRR\ that RR flux can
stabilize closed string tachyons is intruiging.  However in AdS/CFT examples with tachyons from orbifolds (such
as the type 0 theory) one finds a remaining Coleman-Weinberg instability at small radius (where the RR flux
energy density is large and could in principle have removed the instability).  The instability takes the system
to a new vacuum with fewer degrees of freedom, as expected from the worldsheet RG picture of tachyon
condensation, reducing the system to pure Yang-Mills theory at low energies \AdStach\tongtach.

\subsec{Orbifolding to fix moduli}

Another useful technique to fix moduli is orbifolding.  As discussed in the simple example in the last
subsection, an orbifold projects out states which are not invariant under the orbifold action.  This can
eliminate moduli and tachyons.  However it also entails the addition of the twisted sectors, which can
reintroduce moduli and tachyons.

A particularly useful technique for eliminating moduli is {\it asymmetric} orbifolding \asymmetricorb.  This can
be viewed as modding out by T-duality.  For example, let us consider the T-duality action which inverts the
volume of a product of circles.  For each $S^1$, at a generic radius this T-duality action maps one model into
another, but at the self-dual radius T-duality fixes the background.  At this radius, one has zero mode
worldsheet momenta
\eqn\selfdual{\alpha_0={(n+w)\over\sqrt{\alpha^\prime}} ~~~~~ \tilde\alpha_0={(n-w)\over\sqrt{\alpha^\prime}}}
in terms of momentum and winding quantum numbers $n$ and $w$. T-duality acts by taking
$\alpha\to-\alpha,\tilde\alpha\to\tilde\alpha$, or in other words $X_L\to -X_L,X_R\to-X_R$ where $X_L$ and $X_R$
are the left and right moving worldsheet scalars corresponding to the circle direction in the target spacetime.
It leaves the spectrum invariant, as seen by exchanging $n\leftrightarrow w$.

This action has the great virtue of removing the untwisted moduli, since it acts on the corresponding vertex
operators as
\eqn\asymkill{T: (V\sim \del X_L^M\bar\del X_R^N)\to -V}
It is not quite as easy as this makes it sound: there are important constraints on such models, encoded by
modular invariance, and as mentioned above the twisted sectors can reintroduce moduli (and/or tachyons in some
cases).  Solutions to all these constraints preserving the beneficial effects of \asymkill\ have been studied in
\MSS, which we will explore more fully below.

Note that this asymmetric orbifold action is non-geometrical, and one can (and should) explore further
generalizations of string compactifications to include non-geometrical ones, as done for example in the recent
works \nongeom.  In any case, the ingredients and techniques we have covered here so far will suffice to provide
interesting models with compensating forces on all the moduli.

\newsec{Models of Moduli Stabilization}

Using the ingredients catalogued in \S2, it is straightforward to arrange them to provide compensating forces on
the moduli.  In this section we will study three illustrative cases.  For each, we put together appropriate
combinations of the terms and techniques discussed above, and explain the origin of the large numbers required
to obtain a solution expanding around weak coupling or large radius.  The second case here is new and raises
some interesting questions. The third case is the dS models \MSS\kklt.  There have been many other interesting
papers on how to stabilize moduli in other classes of models \achar\Heterotic.

\subsec{Compactification on $S^5$}

Let us warm up by discussing the compactification of type IIB string theory on $S^5$.  This is probably familiar
to you in your studies of the AdS/CFT correspondence, where the corresponding solution is usually presented in
its full ten-dimensional glory.  Here we will simply recover the $AdS_5$ solution using the five-dimensional low
energy effective theory description analogous to that developed in the previous section for the four-dimensional
case.  This provides a useful check on our understanding, and gives a method which generalizes to the cases of
interest such as compactification on Calabi-Yau manifolds, where a full 10d geometry is not known but the low
energy effective field theory is sufficient to determine the low energy potential for the moduli.  (In fact, we
will find in this subsection that the $S^5$ is stabilized at the same radius as that of the $AdS$, which means
the Kaluza-Klein modes need to be included in studying the dynamics of the model.)

Compactifying on an $S^5$ with radius $Ll_s$ produces a negative contribution to the 5d cosmological term from
the positive curvature of the $S^5$.  In string frame, this is
\eqn\adscftR{\Lambda_{\cal R}^{(s)}\sim -{L^5\over e^{2\Phi}}{1\over L^2l_s^2}}
using the fact that the curvature scales like $1/(L^2l_s^2)$, appears at tree level in the $g_s$ expansion, and
is integrated over the whole $S^5$.  In the {\it five}-dimensional Einstein frame, this becomes
 \eqn\adscftRE{\Lambda_{\cal R}^{(E)}\sim -{e^{{4\over 3}\Phi}\over l_5^2L^{16/3}}}
As it must, this contribution goes to zero as the radius $L$ of the $S^5$ goes to infinity.

As it stands, this contribution alone would drive the system toward higher curvature (a big crunch in the $S^5$
directions).  However we can provide a compensating force using the five form flux of the IIB theory.  Putting
$N$ units of this flux on the $S^5$ (and correspondingly on the remaining 5 directions by the self-duality of
the flux in IIB), we obtain a second contribution to the 5d effective theory's potential energy.  This is in
string frame
\eqn\FadsS{\Lambda_F^{(s)}\sim {L^5\over l_s^2} {N^2\over L^{10}}}
and in Einstein frame
\eqn\FadsE{\Lambda_F^{(E)}\sim {N^2\over l_5^2}{e^{10\Phi/3}\over L^{40/3}}}

For $L\to 0$, the flux term dominates while for $L\to\infty$, the curvature term dominates.  If we choose a
sufficiently large coefficient $N$, we can arrange that the two balance at a value of $L$ which is large enough
that this low energy effective field theory analysis (using supergravity alone rather than the full string
theory) applies. At the minimum, the 5d cosmological constant is negative, producing an $AdS_5$ solution.  To
simplify the picture let us define a variable
\eqn\simplvar{\eta\equiv {e^{\Phi/3}\over L^{4/3}}}
In terms of $\eta$, the form of the potential is
\eqn\finalform{\Lambda_E(\eta)=-\eta^4+N^2\eta^{10}}
At the minimum, this satisfies
\eqn\mineta{\eta_*^6\propto {1\over N^2}}
From our definition of $\eta$ \simplvar, we can rewrite this in terms of the original variables as ${L^8\over
g_s^2}\sim N^2$, i.e. the familiar result $L\sim (g_sN)^{1/4}$.  So far this reproduces the expected radius of
the $S^5$ (and the fact that there is a family of solutions obtained by varying $g_s$).  What about the
curvature radius of the $AdS_5$?  In our 5d effective field theory description, this is determined by the value
of the cosmological constant at the minimum.  This is
\eqn\lammin{\Lambda_{min}\sim -{1\over l_5^2}{g_s^{4/3}\over L^{10/3}}{1\over L^2}}
Now using the fact that the 5d Planck length $l_5$ is determined by $l_5^3=g_s^2l_s^3/L^5$, or
$l_5^2=g_s^{4/3}l_s^2/L^{10/3}$, we obtain from \lammin\ that
\eqn\lamminII{\Lambda_{min}\sim {1\over L^2l_s^2}}
So the curvature radius of the $AdS_5$ is indeed $L l_s$.

\subsec{Type IIB on $S^3\times S^3$ With Flux}

Next let me mention a new example, which shares some features with the one we just considered as well as with
the $(A)dS_4$ flux compactifications we will consider later.\foot{I thank S. Hellerman for very useful
discussions on this construction, in particular for alerting me to an important constraint coming from the IIB
axion.} Consider type IIB string theory compactified on $S^3\times S^3$, putting $Q_1,Q_2$ units of RR 3-form
flux $\int F$ on the first and second $S^3$ factors respectively, and similarly $N_1,N_2$ units of NS 3-form
flux $\int H$.  In fact, for simplicity let us take $|N_1|=|N_2|\equiv N$ and $|Q_1|=|Q_2|\equiv Q$, with the
signs determined by a simplifying assumption of cancelling the axion tadpole in \FH.

Using the methods developed in \S2, we will solve the equations of motion for the dilaton and the radii
$R_1l_s,R_2l_s$ of the two 3-spheres (with a round metric for simplicity). First we must take into account the
effects of the $F$ and $H$ flux on other degrees of freedom of the system, as discussed in \S2.4.  In
particular, we can choose the $F$ and $H$ flux to cancel the axion tadpole in \FH. This leaves us with of order
$QN$ anti-D3-branes needed to cancel $\int H\wedge F$ contribution to Gauss' law \gauss.

Now from our considerations in \S2, we can write the structure of the cosmological term in this model:
\eqn\VSIII{\eqalign{\Lambda_E=&{1\over l_4^2}{e^{4\Phi}\over {R_1^6R_2^6}}\biggl(
-{{R_1^3R_2^3}\over{R_1^2}}e^{-2\Phi}-{{R_1^3R_2^3}\over{R_2^2}}e^{-2\Phi}
+{{N_1^2R_2^3}\over{R_1^3}}e^{-2\Phi}+ {{N_2^2R_1^3}\over{R_2^3}}e^{-2\Phi}\cr & +\tilde cNQe^{-\Phi}
+{{Q_1^2R_2^2}\over R_1^3} +{{Q_2^2R_1^2}\over R_2^3}\biggr)\cr}}
The first two terms here come from the positive curvature of the $S^3$s.  The next two come from the NS $H$
flux.  The next term (with order one coefficient $\tilde c$) is the contribution of the order $QN$
anti-D3-branes. Finally, the last two terms are the contributions of the RR fluxes.  The symmetries of the round
sphere metric imply that the equations of motion are automatically solved for the non-radial moduli of the
3-sphere metrics.

It is straightforward to solve $\del \Lambda/\del \Phi = \del \Lambda/\del R_1=\del \Lambda/\del R_2=0$.  The
result is parameterically $e^\Phi\sim N/Q$ and $R_1=R_2\sim N^{1/2}$ in the solution.

Having solved the equations of motion for the moduli, the next step would be to check for tachyons in the
diagonalized mass matrix of $R_1,R_2,\Phi$, as well as the axions and higher KK modes in this space. As far as
the zero mode moduli go, the potential \VSIII\ is of the shape in figure 1 in each direction.  However, as shown
in \garyoliver, the low lying Kaluza Klein modes can sometimes become tachyonic.  I have not completed this
calculation, but it would be interesting to do so along the lines of \garyoliver; our setup is somewhat
different because we do not have a 6-form flux on the compactification, but rather individual 3-form fluxes.

In the next section, we will review among other things the KKLT models with NS and RR flux on 3-cycles of
Calabi-Yau compactifications \kklt.  The model in this subsection may be a good toy model for some aspects of
\kklt, since it involves fluxes on dual three-cycles.  In the $S^3\times S^3$ case, the system might arise via a
finely tuned near horizon limit of D5 and NS5 branes wrapping 3-cycles in the base of a cone over $S^3\times
S^3$.  However, this model, like the $AdS_5\times S^5$ model, has an $AdS$ curvature radius of the same order as
the radii of the $S^3$s.  One can avoid this by orbifolding the sphere(s) by a large group \AdSorb\AcharyaII,
reducing their volume relative to that of the $AdS$ factor or by compactifying on more complicated spaces and
using the three independent terms depicted in figure 2 to stabilize the moduli.  It is to this that we will turn
next.

\subsec{(A)dS Flux Compactifications}

We can now put together our ingredients to stabilize the moduli in more complicated compactifications, such as
realistic ones down to four dimensions with positive vacuum energy, in which the internal compactification can
be parameterically smaller than the curvature radius of the remaining dimensions. I will start by explaining how
the dilaton is stabilized in all known examples, and then discuss how the volume and other moduli are addressed
in the current proposals.

As far as the dilaton goes, we can write its potential in the general form
\eqn\dilgen{\Lambda_E={e^{2\Phi}\over l_4^2}(a-be^\Phi+ce^{2\Phi}+\dots)}
where the $+\dots$ refers to terms subleading and non-perturbative in the expansion around small $g_s=e^\Phi$.
Here the coefficients $a$, $b$, and $c$ can come from the following sources from our list developed above.
First, the coefficient $a$ arises at closed string tree level, and so could get contributions from
dimensionality ($D\ne D_{crit}$), curvature, and NS flux $H^2$ for example.  The $b$ coefficient comes from
orientifolds and D-branes, while the $c$ coefficient comes from RR flux $F^2$ and the 1-loop correction to the
cosmological term.

As far as the quantum contribution to the $c$ coefficient, it is useful to use the Bousso-Polchinski
prescription \BP\ to arrange the RR fluxes to dominate over the 1-loop cosmological term and tune the sum to
close to a desired value.  In order to calculate precisely the fluxes involved in a given model, it would be
necessary to compute this contribution and specify the fluxes required to make such a tune.

In order to stabilize the dilaton at a negative vacuum energy, we can play off a leading order negative term
against a subleading positive term (with a large coefficient to obtain control).  For example, one could
consider a subcritical dimensionality (or positive curvature) to obtain a negative tree-level cosmological term,
balanced against a positive contribution from D-branes or RR flux.  (We saw one version of this in the last
subsection.)

By allowing all three terms in \dilgen\ to play a role, we can produce an interesting ``discretuum" of (A)dS
solutions:  the cosmological constant is discretely variable by varying the dimensionality and the brane and
flux quantum numbers appearing in $a,b$, and $c$.

For example, we can consider a supercritical dimension $D$, orientifolds, and RR fluxes $Q_i$ on the cycles of a
compactification \MSS.  This yields a metastable minimum at which the positive $a$ term, negative $b$ term, and
positive $c$ term balance.  For supercritical dimensions, a toroidal orientifold yields a total contribution
from the orientifolds which scales like $2^{D/4}$, providing a way, by mildly increasing the dimensionality, to
produce a $b$ coefficient in \dilgen\ parameterically larger than the $a$ coefficient. One has of order $2^D$ RR
fluxes to choose from, allowing a Bousso-Polchinski tuning mechanism.

The bare string coupling is fixed in the metastable minimum of \dilgen\ at order $2^{-D/4}$.  However, there are
many species of fields in a generic model of this sort--of order $2^D$ RR fields can propagate in loops in
diagrams. On the other hand, as a function of dimension loop momentum integrals are suppressed by powers of
$1/\Gamma(D/2)$ as explored in \andydim.  It is an open issue to determine the true expansion parameter in the
model and ensure that the calculations are under control. This is a general issue in any model once all moduli
become fixed--it is no longer possible to tune the coupling below any desired value holding other quantities
fixed, and there is always a danger that the true expansion parameter is out of control.

In this model, one can fix all the other NS moduli such as the overall volume via an asymmetric orbifold \MSS.
As discussed above, there are important constraints from modular invariance on asymmetric orbifolding, which for
example allow for the standard GSO projection only in certain dimensions.  However the problem of avoiding
twisted tachyons and moduli becomes easier in higher dimensions, since twist operators in the orbifold CFT have
higher dimensions as we scale up the dimensionality of the target space acted on by the group.

Within the critical string theory, one can similarly fix the dilaton (as well as complex structure moduli of
Calabi-Yau compactifications) via a similar balance of forces \gkpetc.  In this case, to obtain a positive $a$
coefficient in \dilgen, one includes NS flux $H^2$.  Again one considers orientifold 3-planes and RR fluxes to
make up the negative $b$ and positive $c$ terms in \dilgen.  This setup has been developed very elegantly using
the supersymmetric effective field theory packaging into superpotential and Kahler potential terms in
\GVW\gkpetc. Since this has been covered elsewhere, we will here confine ourselves to the component description
developed in these lectures.

In this framework, one can further fix the Kahler moduli by including non-perturbative effects \kklt.  In
particular, there are many models with non-perturbative contributions to the superpotential, scaling like
$e^{-c_{np}L_{cy}^4/g_s}$ from gauge theory effects on 7-branes wrapped on 4-cycles (where $L_{cy}$ is the
length scale of the Calabi-Yau in string units, taken isotropic for simplicity). For example, given a gauge
theory with no massless flavors in four dimensions, coming from wrapped 7-branes, one obtains a gaugino
condensation-induced superpotential.  In the presence of a generic superpotential, one finds $AdS_4$ solutions
preserving supersymmetry.  Adding an anti-D3-brane \kklt\renata\KPV, or starting from a higher energy metastable
vacuum of the flux superpotential \alexetal, produces an independent positive contribution to the potential for
the volume modulus which can dominate at large $L_{CY}$ and play off the non-perturbative effects, producing
very plausibly de Sitter metastable minima of the potential.  In this case, the large numbers introduced include
flux quantum numbers, and relatedly warping of the compactification which allows the antibrane to contributed a
naturally small contribution to the potential which can play off against non-perturbative effects.

Again, once all moduli are fixed the question remains of the expansion parameters involved near large radius and
weak coupling.  As before, the existence of forces playing off each other with the right relative signs to
produce metastable minima, combined with the large number of flux choices available, strongly suggests by
genericity that (A)dS minima obtained by these arguments are available as solutions of the full system.
Nonetheless, it would be useful to produce completely explicit examples in which all relevant effects are
computed, and in which the expansion parameter is determined, at least near the order in which one works, via
explicit computations of the next to leading order effects.

The framework \MSS\ is currently more amenable to a worldsheet string theory analysis, as all contributions
involved in fixing the moduli are perturbative, while the framework \kklt\ is currently more closely tied to
well understood supersymmetric backgrounds of M theory, and is the only existing proposal with low energy
supersymmetry built in, potentially a phenomenologically important feature. Given the genericity of a nontrivial
potential energy with multiple minima in low energy effective field theory, it seems likely that there are many
classes of models, including ones starting from other limits of the theory such as the heterotic string
\Heterotic.

One application of these models is to phenomenology.  One aspect of this is the question of understanding the
distribution of masses and couplings available in low energy limits of string theory.  Another aspect is the use
of the specific features of this space of string vacua for model-building, without attempting to obtain a
general prediction of string theory from the whole set.

Another application is to the physics of inflation, and dark energy.  On the latter subject, one would in
particular like to know the form of holography in four dimensional AdS (and possibly dS) backgrounds. It is to
this that we will turn in the next section.  First we provide an intuitive argument for the number of vacua
expected from this sort of construction.

\subsec{Number of vacua}

The Bousso Polchinski mechanism predicts exponentially many vacua as a function of multiple input flux quantum
numbers, as follows \BP\SKunp\DS\Douglas. The basic idea is the following. One expects a limit on the strength
of flux quantum numbers from back reaction on the geometry. Let us consider for example the KKLT case \kklt.
There are $b_3$ RR flux quantum numbers $Q_i,i=1,\dots,b_3$ and $b_3$ NS flux quantum numbers $N_i, i=1,\dots,
b_3$. If one expresses the expected limitation in the form
\eqn\Rlim{R^2\equiv \sum_{i=1}^{b_3} \gamma_iQ_i^2+ \alpha_iN_i^2 < R^2_{max}}
for some order one coefficients $\alpha_i$ and $\gamma_i$, then one obtains a total number of vacua which is of
order
\eqn\totvac{N_{vac}\sim {R_{max}^{2b_3}\over b_3!}}
from the volume of the sphere in flux space containing the fluxes consistent with \Rlim.  (This assumes that
each choice of flux leads to of order one vacua.)

In the KKLT models, this estimate may be given in terms of the quadratic form
\eqn\Ldef{Q\odot N\sim \int_{CY} H\wedge F}
as follows. Dimensional reduction on a space with flux produces contributions to the four dimensional effective
potential from the flux kinetic terms for the NS flux $H_{NS}$ and the Ramond flux $F_{RR}$
\eqn\fluxkin{\Lambda_{flux}\sim \int_{CY}{1\over l_4^2}{e^{4\Phi}\over V^2}
\sqrt{g}(|F_{RR}|^2+e^{-2\Phi}|H_{NS}|^2).}
where we are in 4d Einstein frame and $V$ is the compactification volume in string units. This contribution
takes the form
\eqn\BPquad{ \Lambda_{flux}\sim \sum_{i=1}^{b_3}(c_iQ_i^2+a_iN_i^2) }
where $a_i$ and $c_i$ are functions of the moduli, which in turn depend on the fluxes, and $Q_i$ and $N_i$ are
the RR and NSNS flux quantum numbers on the 3-cycles in the compactification. (In asymmetric orbifold models
such as \MSS\ the dependence of $a_i,c_i$ on the moduli is eliminated for the geometrical moduli by using
asymmetic orbifolding to freeze them at the string scale.)

If we pick the maximum flux scale $R_{max}$ such that the moduli-dependent coefficients $a_i$ and $c_i$ do not
take extreme values in the solutions to the equations of motion, then one can relate $Q\odot N$ to a positive
definite quadratic form for each point on the moduli space solving the equations of motion.

That is, in the no scale models \GKP\ appearing in KKLT, the Gauss' law relation between $\int H\wedge F$ and
orientifold 3-plane and D3-brane charge
\eqn\gauss{{1\over{2(2\pi)^4(\alpha')^2}}\int H\wedge F = {1\over
4}(N_{O3}-N_{\overline{O3}})-N_{D3}+N_{\overline{D3}}}
translates via supersymmetry into a relation between the orientifold +D3-brane tension and $Q\odot N$.  In a
zero energy vacuum of the no-scale approximation \GKP\ to the effective potential, this tension $\int H\wedge F$
cancels the positive terms \BPquad\ in the potential.  So for every solution to the equations of motion we wish
to consider, a relation of the form
\eqn\boundI{ \sum a_i N_i^2+c_i Q_i^2\sim Q\odot N \le R_{max}^2 }
holds, with $a_i$ and $c_i$ order one coefficients that depend on the fluxes. So rewriting $R_{max}^2$ as
$Q\odot N_{max}$ we can rewrite \totvac\ as
\eqn\Lform{ N_{vac}\sim {(Q\odot N)_{max}^{b_3}\over {b_3!}} }
By integrating the number of vacua solving the equations of motion over the flux choices and moduli space with a
suppression factor introduced for large fluxes to take into account \boundI, \Douglas\ found an estimate
\eqn\Dest{ N_{vac}\sim {(2\pi (Q\odot N)_{max})^{b_3}\over{b_3!}} f(b_3)}
Here $f(b_3)$ is an integral of flux-independent quantities over a fundamental domain of the
moduli space.

This estimate, which may ultimately prove accurate as a count of the number of IIB flux vacua, appears at least
to be a lower bound on this number.  There are other classes of vacua such as \MSS\ to be included in a full
count as well, though the corresponding entropies for these may be studied independently.

\newsec{The Coulomb Branch and Entropy}

In both the AdS and dS case, one expects an entropy associated to the solutions, scaling like
\eqn\entropycc{S\sim {1\over {l_4^2\Lambda}}}
where $\Lambda$ is the Einstein-frame cosmological constant in the (meta-)stable minimum and $l_4$ is the
four-dimensional Planck length.  In the AdS case, this arises from a trivial application of the Susskind-Witten
arguments from AdS/CFT \SW\ to this case, and in the dS case follows from old arguments of Gibbons and Hawking
\GH.  As in black hole physics, this suggests a microphysical count of states, and perhaps a broader holographic
relation.  In this section, I will explain recent progress in extracting these microphysical degrees of freedom
in the flux compactifications to four dimensions.\foot{Other approaches to de Sitter holography were reviewed
recently in \vijayreview.}

Before proceeding, let me dispel a potential confusion. The AdS/CFT duality was argued initially based on near
horizon limits of D-brane systems \juan. In most solutions, even most AdS solutions, such as those developed in
these notes, the solutions are not realized in this way.  The argument based on a near horizon limit involves
extending the system to one with infinitely more degrees of freedom, and then scaling them out to obtain the
precise equivalence discovered in this way by Maldacena \juan.  The AdS/CFT dictionary subsequently articulated
in \AdSdict\ makes no reference to the non-near horizon geometry forming the starting point of the argument
\juan, and indeed the equivalence between AdS string theory and CFT requires no such extension in general. This
strongly motivates proceeding to seek a dual description of the $(A)dS_4$ flux models, but raises at the same
time the question of how to extract the dual description in the absence of a near horizon argument.

One concrete strategy which works is the following.  In ordinary AdS/CFT, if one considers the system on its
Coulomb branch, the gravity side description of the system includes brane domain walls at positions in the
radial direction of the geometry which correspond to the scale of the VEVs turned on in the field theory
description.  In this configuration, the degrees of freedom of the system (e.g. the $N^2$ degrees of freedom of
the ${\cal N}=4$ SYM theory dual to $AdS_5\times S^5$) become manifest on the D-brane domain walls.  This allows
one to view the microphysical degrees of freedom comprising the entropy without ever needing to extend the
system to the non-near-horizon geometry, and without needing to know in advance the field content of the CFT
dual.

This procedure is equally well available also in the $(A)dS_4$ flux models.  For example, in the KKLT geometry
\kklt, the brane domain walls are D5-branes wrapped on cycles dual to those containing the corresponding RR
3-form flux and $NS5$-branes  wrapped on cycles dual to those containing the corresponding NS 3-form flux.  If
the pulled-out branes change $\int H\wedge F$, then there must also be a corresponding number of D3-branes
ending on the domain walls, to satisfy Gauss' law in the compactification for all radial intervals in (A)dS.
This configuration of intersecting 5-branes with 3-branes ending on them is a generalization of the
Hanany-Witten setup \HW\ for studying 3-dimensional gauge theories; our generalization involves live fivebranes
(by which I mean that the degrees of freedom of the 5-branes themselves are dynamical, whereas in the original
Hanany-Witten setup the fivebranes were of infinite extent and produced non-dynamical couplings). It is worth
remarking that in both $AdS_4$ and $dS_4$ flux models, these domain walls are generically time dependent
solutions rather than static configurations.

In \DS, the geometry and causal structure of the Coulomb branch domain wall configurations was determined for
the $dS$ case. In \junctions, the degrees of freedom of the system on the Coulomb branch were identified and
shown to agree with the gravity side prediction, as we review below. In \karch, it was noted that in the de
Sitter case, the system has a ``localized graviton" in addition to the horizon entropy; this makes the dS case
analogous not to pure AdS/CFT but to the AdS/CFT description of a CFT coupled to gravity.

The opposite extreme to string scale geometries is the case where the $(A)dS$ geometry has a curvature radius
tuned to a large value by a generalization of the Bousso-Polchinski mechanism to the flux compactifications. For
these examples, the entropy is predicted to be able to be enormous, in the following way.  In the above
subsection, we found that the number of vacua scales like $N_{vac}\sim {(2\pi (Q\odot
N)_{max})^{b_3}\over{b_3!}} f(b_3)$ \Dest.

If we take these vacua \Dest\ to be distributed roughly uniformly between cosmological constants of $\pm {1\over
l_4^2}$ (where $l_4$ is the four-dimensional Planck length), this predicts a minimum cosmological constant of
magnitude
\eqn\minlam{\Lambda_{min}\sim {1\over {l_4^2N_{vac}}}}
corresponding to a maximum curvature radius $L_{(A)dS}$ of order
\eqn\curvscale{(L^{max}_{(A)dS})^2\sim l_4^2 N_{vac}}
among the elements of the discretuum of vacua predicted by the estimate \Lform\Dest.  This curvature scale in
turn corresponds to an entropy of order
\eqn\entnow{S_{max}\sim {((Q\odot N)^{max}_{(A)dS})^2\over {l_4^2}}\sim N_{vac}}
Taking the vacua to be uniformly distributed is a nontrivial assumption, since the vacua could instead
accumulate around some particular values of the cosmological constant.  We will see that this naive assumption
fits with what we find for the entropy, though a much more thorough analysis of the distribution of vacua will
ultimately be required.

So from the gravity solution, we obtain a prediction for a large entropy, scaling like
\eqn\maxentsimpl{S_{max}\sim {(Q\odot N)_{max}^{b_3}\over b_3!}}

As discussed in \junctions, we can obtain this counting via the degrees of freedom evident on the Coulomb branch
of the solution, by studying the degrees of freedom living on the brane domain walls described above. For
example, we can consider the case where all the (p,q) branes are either D5-branes or NS5-branes (let us say,
D5-branes on all the A-cycles of the Calabi-Yau, and NS5-branes on all the B-cycles), with about the same number
of each.

We have a gauge group $\prod_{i=1}^{b_3}U(N_i)\times U(Q_i)\times U(N_iQ_i)$ from the NS, D5, and D3-branes
respectively. From \boundI\ we see that if we distribute the branes uniformly over the cycles, then $N_i\sim
Q_i\sim \sqrt{Q\odot N/b_3}$. We then have $b_3/2$ bunches of $D_5$ branes, each bunch containing of order
$\sqrt{Q\odot N/b_3}$ branes. Similarly we have $b_3/2$ bunches of wrapped $NS5$-branes, each bunch containing
of order $\sqrt{Q\odot N/b_3}$ branes. Finally we have $b_3/2$ bunches of D3-branes stretching from the $b_3/2$
sets of intersecting A and B cycles, each bunch containing of order $Q\odot N/b_3$ branes.

The degrees of freedom living on the branes consist of string junctions and webs.  If we consider those ending
once on each bunch of branes, they can be electrically charged under each of the $U(n)$ factors in the gauge
group.  This gives them a chance to survive as we go back to the origin of the approximate Coulomb branch,
unlike the (p,q) junctions describing the monopoles and dyons of the ${\cal N}=4$ SYM in the $AdS_5\times S^5$
case.  On the other hand, although our counting will work and makes sense from this point of view, we will not
be able to establish for certain the stability of these degrees of freedom as we go back to the origin.

Counting these junctions, we obtain a maximum entropy
\eqn\maxentII{S_{max}\sim \biggl(\sqrt{Q\odot N\over b_3}\biggr)^{b_3/2}\times \biggl(\sqrt{Q\odot N\over
b_3}\biggr)^{b_3/2} \times \biggl({{Q\odot N}\over b_3}\biggr)^{b_3/2}\sim \biggl({{Q\odot N}\over
b_3}\biggr)^{b_3}}
as predicted in \maxentsimpl.

There are many interesting open questions to pursue in this direction.  For example, we would like to understand
in which precise vacua the junction states are stable as one goes back to the origin.  This is related to the
corresponding question on the gravity side--in which precise flux vacua does the tuning providing a small
cosmological constant arise.

More importantly, we would like to understand the form that holography takes in the (A)dS flux vacua, building
from our nascent understanding of these vacua and their entropy.  The procedure of going on the Coulomb branch,
either to a generic Coulomb branch configuration or via a single brane probe, provides a way to study not just
the microphysical entropy but also some aspects of the dynamics of the putative dual theory--it is an area of
active research.  The form holography might take in dS is a conceptually challenging subject (see e.g.
\karch\dSObj\ for some general discussion of issues involved).  The holographic relation in this case may not be
an exact equivalence to an ordinary non-gravitational system in one lower dimension. However, even at an
approximate level (perhaps at times shorter than the decay time out of a given vacuum) it would be very
interesting to determine the way the degrees of freedom making up the entropy \maxentsimpl\maxentII\ account for
other aspects of the physics of the system.

More generally, the dynamics on the moduli space of interacting scalar field theory has some important
novelties, related to the above issues and to cosmology with scalar field theory.  Let us now turn to this
subject.

\newsec{Velocity Dependent Effects on Moduli Dynamics}

In the preceding sections, I focused on the physics of the scalar potential.  It has become increasingly clear
that the dynamics of scalar fields is significantly affected also by kinetic effects.  Therefore for
completeness I will summarize in this section recent developments involving these effects \Dccel\trapping\
(closely related to earlier developments in the subject of reheating after inflation \kls, earlier works on time
dependent string moduli and D-branes, referenced in \trapping, and earlier comments on the AdS/CFT
correspondence in \KL).

Let us warm up by considering quantum field theory (no gravity) in situations with a moduli space of vacua, such
as the ${\cal N}=4$ super Yang-Mills theory. In interacting field theories, generically there are points in the
moduli space of a scalar field $\phi$ at which light fields $\chi_I$ appear, arising from an interaction of the
form
\eqn\trapint{g^2\sum_{I=1}^{N_\chi}\phi^2\chi_I^2.}
This interaction produces a $\phi$-dependent mass for $\chi$, $m_\chi=g\phi$. When $\phi$ rolls around its
moduli space, this results in a time dependent mass for the $\chi$ particles. When $\phi$ rolls toward $\phi=0$,
the $\chi$ particles become relevant to the dynamics of $\phi$.

This can in general lead to two different effects:  production of $\chi$ particles, and radiative corrections to
$\phi$'s effective action due to $\chi$ particles running in loops.  The first effect, particle production, is
controlled by a non-adiabaticity parameter $\dot m_\chi/m_\chi^2\sim \dot\phi/(g\phi^2)$; the effect is roughly
exponentially suppressed in the inverse of this parameter. The second effect, loop corrections to $\phi$'s
effective action, produce for example higher derivative corrections scaling as $N_\chi \dot\phi^2/\phi^4$. Thus
at weak coupling, we expect particle production to dominate, and at strong 'tHooft coupling (large $gN_\chi$),
we expect the virtual corrections to dominate.  This is indeed what happens, and in both limits the effects of
the $\chi$s back react in a simple but crucial way on the dynamics of $\phi$.

At weak coupling, when $\phi$ rolls past the point $\phi=0$ (say with an impact parameter $\mu$) one finds a
production of the following number density of $\chi$ particles:
\eqn\numdens{ n_{{}_{\chi}} = \int {d^3k\over (2\pi)^3}  n_k = {(gv)^{3/2}\over{(2\pi)^3}} e^{-\pi g \mu^2/v} }
and a corresponding energy density
\eqn\endens{\rho_\chi\sim n_\chi g|\phi(t)|}
This energy density \endens\ traps $\phi$ near the origin, as discussed in detail in \trapping.

In the presence of gravity, the range over which $\phi$'s kinetic energy dominates is limited by Hubble
friction, and the trapping effect can be limited by dilution of created particles.  However, there remains a
nontrivial regime in which this effect persists and provides a vacuum selection mechanism, as well as a
``trapped inflation" mechanism in which the field can be held temporarily on the side of a steep potential,
producing accelerated expansion of the universe.

At strong 'tHooft coupling, one can use the AdS/CFT correspondence to analyze the physics \juan.  The Coulomb
branch of the theory (which in general is not a flat direction) is described on the gravity side by brane domain
walls in $AdS$ (which in general are time dependent solutions moving toward the horizon of the Poincare patch).
On the gravity side, it is immediately clear from the geometry that it takes forever in the (super)gravity
approximation for the domain wall to reach the origin.  Up to back reaction limitations discussed in detail in
\Dccel, the dynamics of the scalar field describing the probe position is governed by the Dirac-Born-Infeld
action, proportional to $1/\gamma \equiv 1/\sqrt{1-v_p^2}$ where $v_p^2=N \dot\phi^2/\phi^4$ is the proper
velocity of the probe.  This leads to a dynamics in which the higher derivative terms in the action contribute
crucially, slowing down the field velocity $\dot\phi$ proportionally to $\phi^2$ as it approaches the origin, an
effect dubbed ``D-cceleration".

When coupled to gravity by embedding the system in a Calabi-Yau compactification, there are generically mass
terms generated for $\phi$ and a corresponding closing up of the $AdS$ throat at a finite nonzero warp factor in
the IR \kklmmt.  This preserves a window in which the D-cceleration effect operates and produces inflation on a
steep potential.  The density perturbation spectrum, obtained by expanding the DBI action coupled to gravity,
leads to distinctive predictions of this inflationary model coming from controlled but relatively large
non-Gaussian contributions \Dccel.

I include this brief review of velocity dependent effects because, aside from their intrinsic interest, it is
important to account for them in determining the dynamics of the moduli.  Putting together what we have learned
here about the moduli potential and kinetic effects, and their relation to the microphysics of string theory, I
hope you will be better prepared to attack the many important open problems that await us.

\noindent{\bf Acknowledgements}

I would like to thank the organizers and participants of the PiPT, TASI, and ISS schools for stimulating
programs. I am grateful to many wonderful collaborators, as well as to many other colleagues working in these
areas for useful discussions.  I would like to thank D. Tong additionally for comments on the manuscript. This
research is supported in part by the DOE under contract DE-AC03-76SF00515 and by the NSF under contract 9870115.

\listrefs

\end